\documentclass[aps,prb,reprint]{revtex4-1}

\usepackage{amsmath,amsfonts,amssymb}
\usepackage{graphicx}
\usepackage{subfigure}

\begin{document}

\title{Understanding spintronics in F/N/F structures through a mechanical analogy}

\author{Ya.\ B. Bazaliy}
\email{bazaliy@mailbox.sc.edu}
\affiliation{Department of Physics and Astronomy, University of South Carolina,
 Columbia, SC 29208}

\date{\today}

\begin{abstract}
A mechanical equivalent system is introduced to mimic the behavior of multilayer structures with diffusive spin transport. The analogy allows one to use existing mechanical intuition to predict the influence of various parameters on spin torques and spin-dependent magnetoresistance. In particular, it provides an understanding of the sign-changing behavior of spin torque in asymmetric F/N/F spin valves. It further helps to uncover the physical reason behind the singular behavior of spin magnetoresistance in devices with ultra-thin N-layers.
\end{abstract}

\maketitle

\section{Introduction}
Electric current flowing through a magnetic structure with spatially non-uniform magnetization ${\bf M}({\bf r})$ is known to produce torques that grow linearly with current magnitude. These torques were originally understood in terms of spin transport and called spin-transfer torques.\cite{berger,slon_1996} They can be equivalently understood as being produced by the exchange interaction $J ({\bf s} \cdot {\bf M})$ between the spins of itinerary electrons $\bf s$ and the magnetization.

\begin{figure}[b]
  \centering
  \includegraphics[width=0.4\textwidth]{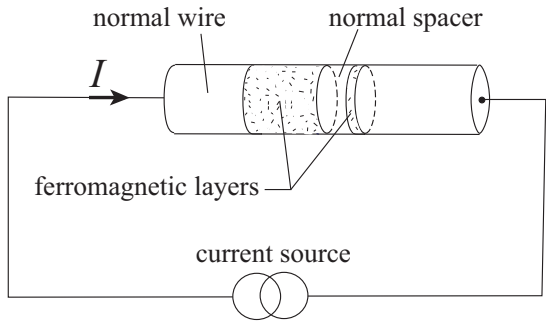} \\
  \caption{Spin-transfer device.}
  \label{fig:device}
\end{figure}

Exchange interaction produces a torque ${\bf T} = J [{\bf M} \times \langle {\bf s} \rangle ]$ acting on the magnetization, where $\langle {\bf s} \rangle$ is the spin density of itinerant electrons. Imagine a ferromagnet with non-uniform magnetization. In equilibrium $\langle {\bf s}({\bf r}) \rangle$ is parallel to ${\bf M}({\bf r})$ everywhere in the sample, so ${\bf T} = 0$. In the presence of electric current the situation changes. Electrons arriving to a point ${\bf r}$ with the current flow bring their spins from afar, where they were directed at some angle to ${\bf M}({\bf r})$. A non-equilibrium state with non-collinear vectors $\langle {\bf s} \rangle$ and ${\bf M}$ is thus formed, and ${\bf T}$ acquires a non-zero value. It's magnitude grows with increasing departure from equilibrium, i.e., with the magnitude of the driving electric current.

The same physics can be equvalently described in terms of the transport of angular momentum $\hbar/2$, associated with each itinerary electron.\cite{berger, slon_1996, dugaev:2006, ryabchenko:2010} In this approach  the torque is associated with spin currents ${\bf j}_s$. Consider a physically infinitesimal element of a ferromagnet. The spin denisty $\langle {\bf s} \rangle$ of itinerary electrons inside this element can change for two reasons: (a) exchange interaction produces a torque ${\bf T}_s$ acting on $\langle {\bf s} \rangle$; (b) incoming and outgoing spin currents do not compensate each other, so $\langle {\bf s} \rangle$ changes due to electron motion. In a stationary state $\langle {\bf s} \rangle = {\rm const}$, so, the term produced by exchange must be exactly compensated by the term arising from the spin currents imbalance. In continuous medium description the  latter is given by the divergence of spin current, and ${\bf T}_s$ is found to be proportional to it. Finally, since the torque $\bf T$ acting on ${\bf M}$ and the torque ${\bf T}_s$ acting $\langle {\bf s} \rangle$ are the results of one and the same interaction, one has ${\bf T} = -{\bf T}_s$ (action equals counter-action). To sum up, in a stationary state the torque ${\bf T}({\bf r})$ is determined by the divergence of spin current at the observation point ${\bf r}$.

In this work we will discuss multilayers formed by ferromagnetic (F) and normal (N) metals. In those structures $\bf M$ changes discontinuously at the boundaries between materials. Experimentally spin-transfer torque is often observed in devices consisting of a normal metal wires with diameters of the order of $100$ nm, having two ferromagnetic inclusions (Fig.~\ref{fig:device}). If the magnetization directions of the F-layers are given by the unit vectors ${\bf m}_1$ and ${\bf m}_2$, one finds \cite{slon_1996, slon:2002, stiles:2002} that the torques acting on ${\bf M}_1$ and ${\bf M}_2$ are given by the expressions
\begin{eqnarray}
\nonumber
 {\bf T}_1 &=& -\frac{I\hbar}{2 e} \ g_1(\theta)  [{\bf m}_1 \times [{\bf m}_2 \times {\bf m}_1]] \ ,
\\
\label{eq:T12}
 {\bf T}_2 &=& \frac{I\hbar}{2 e} \  g_2(\theta)  [{\bf m}_2 \times [{\bf m}_1 \times {\bf m}_2]] \ .
\end{eqnarray}
Here $I$ is the electric current in the wire, $e < 0$ is the electron charge (so that  $I/e = j_0$ is the particle current), $\theta$ is the angle between ${\bf m}_1$ and ${\bf m}_2$, and   $g_{1,2}(\theta)$ are the ``efficiency factors''. The opposite signs in the torque formulas reflect the symmetry of the problem: namely, the F-layer located downstream for a positive current will be in an upstream position for a negative current. Note that although the two torques are produced by an electron-mediated interaction between the F-layers, their sum is non-zero, ${\bf T}_1 + {\bf T}_2 \neq 0$. This does not violate the angular momentum conservation because the ferromagnetic layers do not form a closed system and can transfer angular momentum to the other parts of the structure where it is damped into the crystal lattice.

Efficiency factors are determined by the character of electron transport, material parameters, and the geometry of the device. Their dependence on the angle $\theta$ was considered in many studies. Early investigators took it as evident that efficiency factors are positive. This view was supported by the following arguments. Electrons entering the normal spacer from the F$_1$ layer remain spin-polarized along ${\bf m}_1$. When they reach the F$_2$ layer, their spins have to rotate to match (on average) the direction of ${\bf m}_2$. The requirement of spin rotation makes it harder for the electrons to cross the boundary with the second ferromagnet, leading to an increase in the device resistance. The larger is the angle, the bigger will be the resistance increase. The flip side of this observation is that the same physical processes that enable the influence of the angle $\theta$ on the resistance of the structure will cause that structure to adjust its magnetic state so as to ease the passage of electrons pumped by the current source.\cite{pipe-analogy} Thus the emerging torque should push ${\bf m}_2$ towards ${\bf m}_1$ to decrease $\theta$. The direction of ${\bf T}_2$ is consistent with this conclusion if $g_2(\theta) > 0$ holds for all values of the angle.

However, subsequent calculations \cite{kovalev:2002, manschot:2004, barnas:2005} had shown that already within the model of diffusive transport the efficiency factor can be a sign-changing function of the angle in sufficiently asymmetric structures with F$_1$ and F$_2$ layers made of different materials. This result was used in numerous proposals of magnetic oscillators powered by dc current.\cite{manschot:2004, gmitra:2006, balaz:2009} Calculations of the efficiency factor are straightforward but cumbersome, so there is a need for a simple and intuitive understanding of the sign-changing behavior of $g(\theta)$. Here we show that such understanding can be achieved through a mechanical analogy, and set up a graphic interpretation method that allows one to obtain qualitative results without performing detailed calculations.

\section{Diffusive transport}
\subsection{Bulk equations}

If the mean free path of electrons is much shorter than the layer thickness in Fig.~\ref{fig:device}, transport can be described by diffusive equations. Assuming that the current is uniform in the wire's cross-section, all physical quantities depend on a single coordinate $x$ along the wire. Ferromagnetic inclusions are characterized by two diffusion coefficients $D_{\uparrow}$ and $D_{\downarrow}$ describing electrons with spins parallel (up) and anti-parallel (down) to the magnetization direction $\bf m$. Normal parts of the wire are described by a single diffusion coefficient $D_N$. In an infinite ferromagnetic wire electric current is accompanied by a spin current ${\bf j}_s = p j_0 {\bf m}$, where $p = (D_{\uparrow} - D_{\downarrow})/(D_{\uparrow} + D_{\downarrow})$ is the transport spin polarization and $j_0$ is the particle current. Since we consider an effectively 1D situation, $j_0$ is a scalar and ${\bf j}_s$ is a vector in spin space.

Electric current produces non-equilibrium spin accumulations $\delta {\bf s}$ near the layers' boundaries.\cite{vanson:1987, valet-fert:1993} In normal layers $ \delta {\bf s} = \langle {\bf s} \rangle$, and in ferromagnetic layers $ \delta {\bf s} = \langle {\bf s}  \rangle - {\bf s}_{eq}$, where ${\bf s}_{eq}$ is the equilibrium spin density. We will consider strong ferromagnets for which the spin band splitting $J$ is of the same order as the Fermi energy $\epsilon_F$. Large $J$ guarantees that  ${\bf s}_{eq}$ and $\delta {\bf s}$ are always collinear with ${\bf m}$.\cite{nazarov:2000, stiles:2002} If a spin current of arbitrary polarization in spin space impinges on the N/F boundary from the N-layer side, then inside F its polarization differs from $\bf m$ only in a boundary layer of thickness $\lambda_J \sim \hbar v_F/J$, which for $J \sim \varepsilon_F$ is of the order of electron wavelength and much smaller than the mean free path. In the diffusive approximation, where the mean free path is considered to be infinitesimally small, such behavior is encoded in the boundary conditions at the N/F interfaces.

We would like to underscore that strong ferromagnets with $J \sim \epsilon_F$ do not necessarily exhibit a perfect transport spin polarization $p = 1$. The latter is determined not only by the value of $J$ but also by the Fermi velocities and scattering times of itinerant electrons. Because of that $p$ may be arbitrarily small in strong ferromagnets. Conversely, if the $\delta {\bf s}, {\bf j}_s || {\bf m}$ requirements have been imposed on the system, considering the $p \to 0$ limit is not automatically equivalent to making a particular layer normal.

It is well known\cite{kovalev:2002, heide:2002} that in the diffusive approximation spin accumulation can be characterized by a splitting of chemical potentials of up- and down-spin electrons, $\mu_s = \mu_{\uparrow} - \mu_{\downarrow}$, with ``up'' and ``down'' defined relative to the local $\delta {\bf s}(r)$.  Both the magnitude and the direction of spin accumulation are encoded in a vector spin potential $\boldsymbol{\mu}_s = \mu_s {\bf m}$, where the unit vector ${\bf m}(r)$ is pointing along $\delta {\bf s}(r)$. In the device shown in Fig.~\ref{fig:device} vector ${\bf m}(r)$ can vary spatially in the N-layers but will point along ${\bf m}_{1}$ and ${\bf m}_{2}$ in the ferromagnetic inclusions.

Particle and spin currents are determined by the gradients of $\boldsymbol{\mu}_s$ and electro-chemical potential  $\mu_0$.\cite{rashba_epjb2002} In ferromagnetic layers
\begin{eqnarray*}
 j_0 &=& -(D_{\uparrow} + D_{\downarrow})\left[
   \nabla_x \mu_0 + p \frac{\nabla_x \mu_s}{2}
 \right] \ ,
 \\
 {\bf j}_s &=& -(D_{\uparrow} + D_{\downarrow})\left[
 p \nabla_x \mu_0 + \frac{\nabla_x \mu_s}{2}
 \right] {\bf m} \ ,
\end{eqnarray*}
while in the normal metal layers
\begin{eqnarray}
 \nonumber
 j_0 &=& - D_{N} \nabla_x \mu_0 \ ,
 \\
 \label{js_N}
 {\bf j}_s &=& - D_N \frac{\nabla_x \boldsymbol{\mu}_s}{2} \ .
\end{eqnarray}
Taking advantage of the constancy of particle current $j_0 =  {\rm const}$ in the 1D situation, one can re-write spin current in ferromagnets as
\begin{equation}\label{js_FM}
{\bf j}_s = \left( p j_0 - D_F \frac{\nabla_x \mu_s}{2} \right){\bf m} \ ,
\end{equation}
with $D_F = 4 D_{\uparrow} D_{\downarrow}/(D_{\uparrow} + D_{\downarrow})$.

The distributions of spin currents and spin accumulations are governed by the diffusion equation with spin relaxation terms \cite{rashba_epjb2002}
$$
\frac{d\boldsymbol{\mu}_s}{dt} + \frac{1}{\rho} \ {\rm div}{\bf j}_s = - \frac{\boldsymbol{\mu}_s}{\tau} \ ,
$$
where $\tau$ is the spin relaxation time in the material, $\rho = ds/d\mu_s$ is a coefficient obtained from the densities of states for two spin directions, and ${\bf j}_s$ is given by Eqs.~(\ref{js_N}) or (\ref{js_FM}). We will consider stationary states, where
\begin{equation}\label{divJ}
\frac{1}{\rho}\frac{d{\bf j}_s}{dx} = - \frac{\boldsymbol{\mu}_s}{\tau} \ .
\end{equation}
As one can see, in a stationary situation equations for spin current decouple from those for particle current.

To find spin currents and spin accumulations everywhere we have to solve Eq.~(\ref{divJ}) with appropriate boundary conditions at the interfaces between the layers.

\subsection{Boundary conditions}
In the case of electric current flowing through a device with material boundaries, the approximation of choice depends on the relative magnitudes of the interface and bulk resistances. If the interface resistances are small, one can assume that electric potential is continuous at the boundaries. This is the transparent boundary approximation, in which there are no jumps of either potential or current.

The case of spin transport is different. Here one expects to find jumps of spin current across the N/F interfaces. Indeed, in F-layers spin current is always polarized along the magnetization but in N-layers there is no such restriction, so it is easy to imagine spin currents having different polarizations on the N and F sides of the interface. After all, in contrast to the case of particle current, a discontinuity of spin current is not forbidden by the conservation laws. Do the jumps of spin current necessarily produce the jumps of spin potential?
Microscopic models \cite{nazarov:2000} prove the existence of spin transparent interfaces, such that both $\mu_0$ and $\boldsymbol{\mu}_s$ potentials are continuous across them. Spin currents have some continuity properties as well. Namely, when the boundary can be modeled by a static, spin-dependent potential, the paral\-lel component of spin current, $({\bf j}_s \cdot {\bf m})$, is found to be continuous. Such boundaries are called ``spin inactive''. The ``inactivity'' refers to the parallel component only. The perpendicular component of spin current is generally discontinuous at the N/F boundaries with strong ferromagnets.

It was shown in Refs.~\onlinecite{kovalev:2002, manschot:2004, barnas:2005} that the boundary conditions read
\begin{eqnarray}
\label{BC-mu}
{\boldsymbol{\mu}_s}|_N &=& {\boldsymbol{\mu}_s}|_F ,
\\
\label{BC-Jparallel}
({\bf j}_s \cdot {\bf m})|_N &=& ({\bf j}_s \cdot {\bf m})|_F \ ,
\end{eqnarray}
and provide enough constraints to find spin current every\-where in the wire. They do not impose any restrictions on the perpendicular components of spin current. Those can be found from the bulk equations.

\begin{figure}[t]
  \centering
  \includegraphics[width=0.22\textwidth]{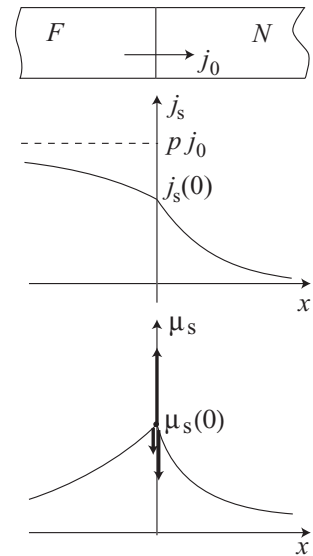} \\
  \caption{Spatial distributions of $j_s(x)$ and $\mu_s(x)$ in an F/N contact. The arrows on the $\mu_s(x)$ graph depict the forces considered in the mechanical analogy.}
  \label{fig:NFcontact}
\end{figure}

\section{Mechanical analogy and graphic approach}

\subsection{F/N contact}
For the purpose of introducing the notions of the suggested mechanical analogy we start with a well known case of a single F/N contact (Fig.~\ref{fig:NFcontact}) with spin-inactive interface and half-infinite F and N layers.\cite{vanson:1987} Let the interface be located at $x = 0$. At the F/N boundary the spin potential is continuous and has a magnitude $\boldsymbol{\mu}_s(0)$. In general, to set up the problem in a given layer one needs to spe\-cify $j_0$ and two spin potentials at the left and right ends of that layer. For the F layer these are $\boldsymbol{\mu}_s(- \infty)$ and $\boldsymbol{\mu}_s(0)$, and for the N layer---the $\boldsymbol{\mu}_s(0)$ and $\boldsymbol{\mu}_s(+ \infty)$. Unlike the particle current, spin current ${\bf j}_s(x)$ is not constant and should be obtained by solving Eq.~(\ref{divJ}). The spin potential $\boldsymbol{\mu}_s(x)$ is non-zero near the interfaces and should vanish far away from them, meaning $\boldsymbol{\mu}_s(\pm \infty) = 0$. In the F/N contact ${\bf j}_s$ and $\boldsymbol{\mu}_s$ are parallel to the F-layer magnetization  everywhere in the structure, $\boldsymbol{\mu}_s(x) = \mu_s(x) {\bf m}$, ${\bf j}_s = j_s(x) {\bf m}$. Solutions of Eq.~(\ref{divJ}) read
\begin{eqnarray*}
 \mu_s(x) &=& \mu_s(0) \exp(x/l_{F}) \qquad (x<0) \ ,
 \\
 \mu_s(x) &=& \mu_s(0) \exp(-x/l_{N}) \qquad (x>0) \ ,
\end{eqnarray*}
where $l = \sqrt{D\tau/2\rho} \ $ denotes the spin diffusion length in the corresponding layer.

The value of $\boldsymbol{\mu}_{s}(0)$ is then obtained from the boundary condition (\ref{BC-Jparallel}). In the present situation it is given by an equation
 \begin{equation} \label{BC-Jparallel_FN}
 p j_0 - G_F \ \mu_s(0) - G_N \ \mu_s(0) = 0 \ .
\end{equation}
with the definition $G = D/2l$ in each layer.

Spatial distributions of spin current and spin potential are shown in Fig.~\ref{fig:NFcontact}. Their shape can be understood through the following mechanical analogy. Imagine the parts of the $\mu_s(x)$ graph for $x > 0$ and $x < 0$ to be elastic cords, attached to the horizontal axis at $ x =\pm \infty$, and to a small weightless ring at $x = 0$. The ring can slide without friction along a vertical rod. An upward force $p j_0$ is applied to the ring. The elasticity of the cords produces two downward forces $G_F \mu_s(0)$ and $G_N \mu_s(0)$ that are proportional to the deviation of the ring from the origin. Then Eq.~(\ref{BC-Jparallel_FN}) can be viewed as a condition of balance of forces acting on the ring. A constant driving force pulls the ring up and away from the origin, and the two restoring forces pull it back. The whole setup is equivalent to a mechanical system with two springs connected in parallel and being expanded by an applied force. Using this mechanical picture one can easily and intuitively predict  the dependence of $\mu_s(0)$ on the system parameters. For example, since the spring coefficients are proportional to $1/l$, and since faster spin relaxation in either of the materials decreases $l$, the springs of the mechanical analogy become stronger with increasing spin relaxation, and the spin accumulation gets smaller. When $\mu_s(0)$ is found, spin current $j_s(x)$ can be obtained everywhere from the formulas (\ref{js_N}) and (\ref{js_FM}).

Note that for an N/F contact, with particles moving from N to F, Eq.~(\ref{BC-Jparallel}) gives
\begin{equation} \label{BC-Jparallel_NF}
 - p j_0 - G_F \ \mu_s(0) - G_N \ \mu_s(0) = 0 \ .
\end{equation}
From the point of view of mechanical analogy the external force now pulls the ring downward.

\begin{figure}[t]   
 \center
  \subfigure{\includegraphics[width = 0.20\textwidth]{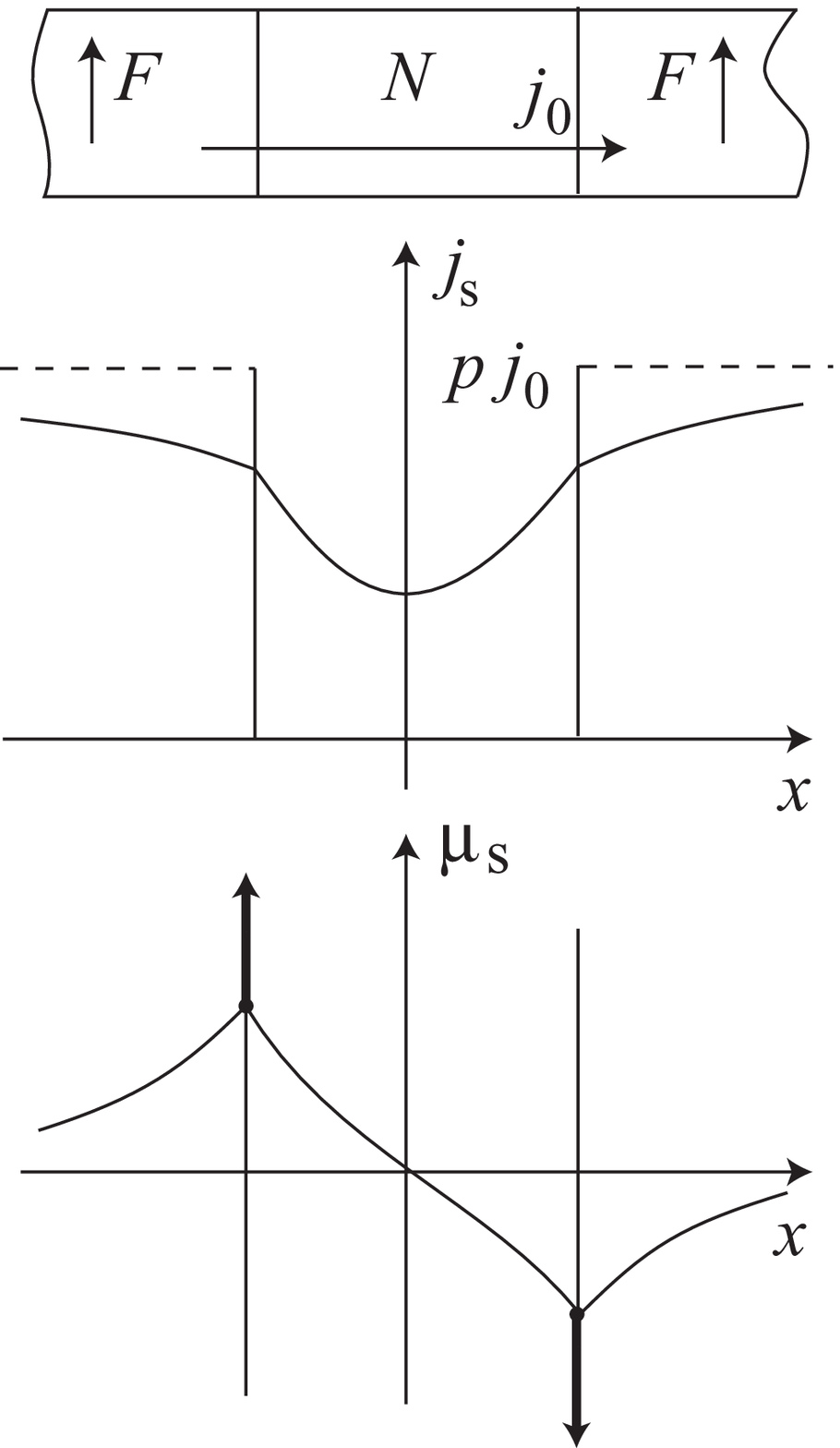}}
  \subfigure{\includegraphics[width = 0.20\textwidth]{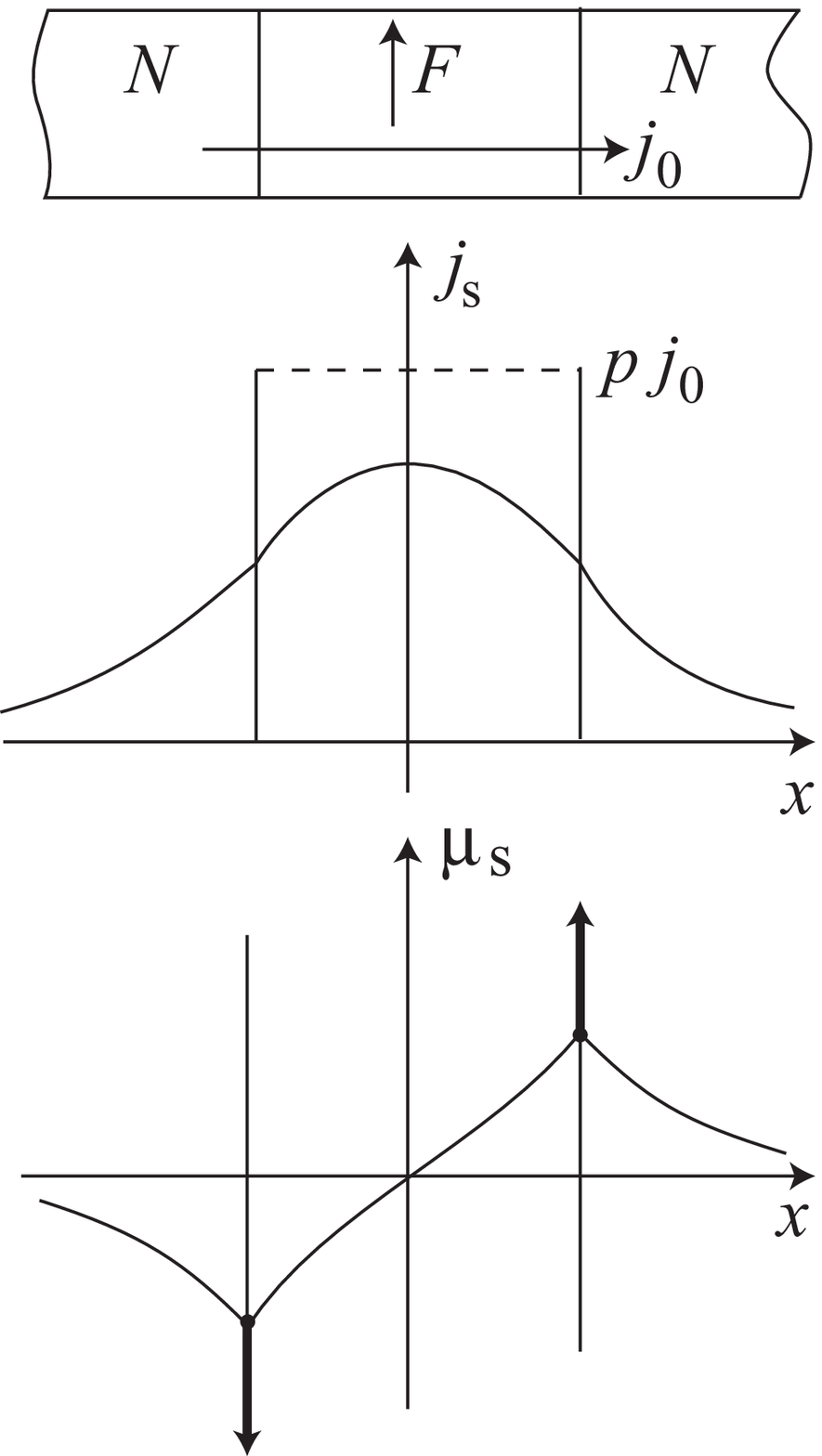}}
\caption{Spatial distributions of $j_s(x)$ and $\mu_s(x)$ in collinear F/N/F and N/F/N structures. The lower panel shows how the directions of forces employed in the mechanical analogy depend on the order of N and F layers along the particle flow direction}
 \label{fig:collinear}
\end{figure}

Using the rules of the mechanical analogy for the F/N and N/F contacts, we can easily expand our understanding to the collinear F/N/F and N/F/N structures (Fig.~\ref{fig:collinear}). Here each boundary is characterized by its own driving force, and the two forces pull the connecting weightless rings in the opposite directions.

\subsection{Non-collinear structure} \label{sec:noncollinear}
\begin{figure}[t]
 \centering
 \includegraphics[width=0.35\textwidth]{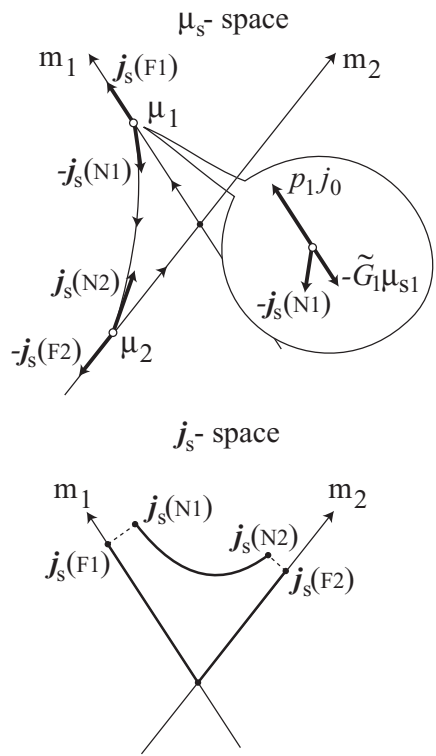}
\caption{Hodographs $\boldsymbol{\mu}_s(x)$ and ${\bf j}_s(x)$.}
  \label{fig:parametric_normal}
\end{figure}

First, consider a non-collinear structure from the traditional point of view based on diffusive transport equations. The simplest structure with a sign-changing efficiency factor $g(\theta)$ is the N/F$_1$/N/F$_2$/N structure with vanishing spin diffusion lengths in the outer normal layers.\cite{kovalev:2002} It is evident from the discussion above that in the limit $l \to 0$ in the outer N-layers the spin potential vanishes at the N/F$_1$ and F$_2$/N interfaces. Spin potentials $\boldsymbol{\mu}_{1}$ at the F$_1$/N interface ($x = x_1$) and $\boldsymbol{\mu}_{2}$ at the N/F$_2$ interface ($x = x_2$) should be derived from the condition (\ref{BC-Jparallel}) applied at those interfaces.
Solutions of the bulk equation (\ref{divJ}) in each layer have a form $\boldsymbol{\mu}_s(x) = {\bf A} \exp(-x/l) +  {\bf B} \exp(x/l)$ with constant vectors $\bf A$ and $\bf B$ determined by $\boldsymbol{\mu}_{s}(L)$ and $\boldsymbol{\mu}_{s}(R)$ at the left and right boundaries of that layer. This means that within one layer $\boldsymbol{\mu}_s(x)$ varies in a plane of spin space defined by
$\boldsymbol{\mu}_{s}(L)$ and $\boldsymbol{\mu}_{s}(R)$.
Matched solutions for the whole structure were found this way in Refs.~\onlinecite{kovalev:2002, barnas:2005} (see Appendix \ref{appendix-I}). Below we will concentrate on understanding them through the mechanical analogy and its appropriate graphical representation. Generally, a vector function $\boldsymbol{\mu}_s(x)$ can be represented by a hodograph in the 3D spin space. But since we know that in a given layer $\boldsymbol{\mu}_s(x)$ belongs to a plane, one layer can be represented by a 2D hodograph. Furthermore, in a structure with just two ferromagnets vector $\boldsymbol{\mu}_s(x)$ always belongs to the plane defined by vectors ${\bf m}_1$ and ${\bf m}_2$. The end of vector $\boldsymbol{\mu}_s(x)$ moves along the ${\bf m}_1$ and ${\bf m}_2$ lines in the ferromagnets, and along a curve connecting $\boldsymbol{\mu}_{1}$ and $\boldsymbol{\mu}_{2}$ in the normal metal (see Fig.~\ref{fig:parametric_normal}, upper panel). Boundary condition (\ref{BC-Jparallel}) applied at the points $x_1$ and $x_2$ lead to the relations
\begin{eqnarray}
\nonumber
&& ({\bf j}_s(F_1) - {\bf j}_s(N_1)) \cdot {\bf m}_1 = 0 \ ,
\\
\label{BC-Jparallel_FNF}
&& ({\bf j}_s(N_2) - {\bf j}_s(F_2)) \cdot {\bf m}_2 = 0 \ ,
\end{eqnarray}
where arguments (F$_1$) and (N$_1$) denote the points $x_1 \mp 0$, while arguments (N$_2$) and (F$_2$) refer to the $x_2 \mp 0$ points.

The mechanical analogy for a non-collinear structure is set up as follows. The hodograph lines in Fig.~\ref{fig:parametric_normal} are replaced by elastic cords. The cord representing the N layer is connected to a pair of small weightless rings at the $\boldsymbol{\mu}_{1}$ and $\boldsymbol{\mu}_{2}$ points. Those rings slide without friction along the two crossed rods, directed along ${\bf m}_1$ and ${\bf m}_2$. It is assumed that each ring can pass through the crossing point of the rods: one can imagine that the rods are slightly parallel-shifted away from each other. The cords representing the F$_{1,2}$ layers are connected to the respective rings at one end, and to the crossing point of the rods at the other. Altogether, the F-layer cords are going straight from the $\boldsymbol{\mu}_{1,2}$ points towards the rods crossing, and the N-layer cord is hanging with a slack between $\boldsymbol{\mu}_{1}$ and $\boldsymbol{\mu}_{2}$. The elastic forces applied by the cords to the rings are mapped on the spin currents at the interfaces. The cord tension is given by the absolute value of spin current. According to Eq.~(\ref{js_N}) vectors ${\bf j}_s$ on the N sides of the interfaces are directed along the tangents to the N-hodograph, just as the cord tension force. Fig.~\ref{fig:parametric_normal} shows the expanded N-cord applying forces $-{\bf j}_s(N_1)$ to the $\boldsymbol{\mu}_{1}$ ring, and ${\bf j}_s(N_2)$ to the $\boldsymbol{\mu}_{2}$ ring. In mechanical terms conditions (\ref{BC-Jparallel_FNF}) establish the balance of forces on the rings. Since the rings are restricted to move along the rods, only the parallel components of the forces are balanced.

Spin current expressions in ferromagnets have two terms, e.g., $j_s(F_1) = p_1 j_0 - \tilde G_1 \mu_1$ (function $\tilde G_1$ is defined in Appendix~\ref{appendix-I}). Similar to the collinear case, the first term, proportional to the particle current, is mapped on a force pulling the ring away from its equilibrium position at the rods crossing point (see inset in Fig.~\ref{fig:parametric_normal}, upper panel). The second term produces an elastic restoring force, pulling the ring back to equilibrium. As expected, the driving forces in the mechanical analogy correspond to the driving terms in the diffusive equations, and are proportional to the magnitude of the current pumped through the device.

Mechanical analogy also applies to Eq.~(\ref{divJ}) that determines the hodograph shape in the N-layer. When re-written as $- d{\bf j}_s/dx - (\rho/\tau)\boldsymbol{\mu}_s = 0$, this equation can be interpreted as a differential form of the balance of forces for an infinitesimal element of the cord. The latter is acted upon by two tension forces applied to its ends, and a distributed force, proportional to the elements length, pointing to the crossing point of the rods. It is this ``attraction to the origin'' force that is responsible for the slack of the N-layer cord shown in Fig.~\ref{fig:parametric_normal}. This force is inversely proportional to $\tau$, so, in the absence of spin relaxation in the normal metal, i.e., $\tau \to \infty$, the cord will go straight between the $\boldsymbol{\mu}_{1}$ and $\boldsymbol{\mu}_{2}$ points.

Note that the stationary transport equations (\ref{divJ}) and the boundary conditions (\ref{BC-mu}, \ref{BC-Jparallel}) can be obtained by searching for a minimum of a functional
$$
S = \sum_{\alpha} \int \left[ \frac{1}{D_{\alpha}} \left( p_{\alpha} j_0 -  D_{\alpha} \frac{\nabla_x \boldsymbol{\mu}_s}{2} \right)^2
 + \frac{\rho_{\alpha}\boldsymbol{\mu}_s^2}{\tau_{\alpha}}
 \right] dx \
$$
in the class of continuous functions $\boldsymbol{\mu}_s(x)$ with a restriction $\boldsymbol{\mu}_s(x) || {\bf m}$ in ferromagnetic regions. In the definition of $S$ the sum goes over all layers indexed with $\alpha$, and the integrations are performed within each layer. Material parameters $p_{\alpha}$, $D_{\alpha}$, $\tau_{\alpha}$ and $\rho_{\alpha}$ a constant within each layer. From the point of view of diffusive transport this result can be viewed as an example of minimal entropy production principle. From the point of view of mechanical analogy the first term in the integrand represents the elastic energy of the strained cord, and the second one captures the energy related to the attraction force.

Overall, the suggested analogy allows one to make qualitative predictions about the solutions of diffusive transport equations by engaging our intuition about the behavior of mechanical systems under the action of external forces. We show that below for the case of spin-transfer torque.

\subsection{Spin-transfer torques and sign-changing efficiency factor}
Spin-transfer torque ${\bf T}$ is determined by the jump $\Delta {\bf j}_s$ of spin current at the N/F interface as ${\bf T} = (\hbar/2) \Delta {\bf j}_s$.\cite{slon_1996} According to the boundary conditions (\ref{BC-Jparallel}) the jumps of ${\bf j}_s$ and $\bf T$ are perpendicular to the magnetization of the F-layer, in agreement with formulas (\ref{eq:T12}). In our N/F$_1$/N/F$_2$/N structure the jumps happen at the F$_1$/N and N/F$_2$ boundaries and are given by the expressions
\begin{eqnarray}
\nonumber
\Delta {\bf j}_{s1} &=& {\bf j}_s(F_1) - {\bf j}_s(N_1)
 = j_s(F_1) {\bf m}_1 - {\bf j}_s(N_1) \ ,
 \\
 \label{eq:js_jumps}
\Delta {\bf j}_{s2} &=& {\bf j}_s(N_2) - {\bf j}_s(F_2) =
 {\bf j}_s(N_2) - j_s(F_2)  {\bf m}_2 \ .
\end{eqnarray}

The hodograph of vector  ${\bf j}_s(x)$ is shown in Fig.~\ref{fig:parametric_normal} (lower panel). Because of the jumps of spin current, it comprises three disconnected segments. The N-layer segment goes between the ${\bf j}_s(N_1)$ and ${\bf j}_s(N_2)$ points. The F-layer segments go from the origin to the ${\bf j}_s(F_1)$ and ${\bf j}_s(F_2)$ points. Solutions of the diffusive equations show that the stronger is spin relaxation in N, the more pronounced is the slack of the ${\bf j}_s$ hodograph, and the smaller are the spin current jumps at the interfaces. In the opposite limit of vanishing spin relaxation the hodograph of $\boldsymbol{\mu}_s(x)$ becomes a straight line, and the hodograph of ${\bf j}_s(x)$ shrinks to a single point (spin current is conserved).

\begin{figure}[t]
  \centering
  \includegraphics[width=0.35 \textwidth]{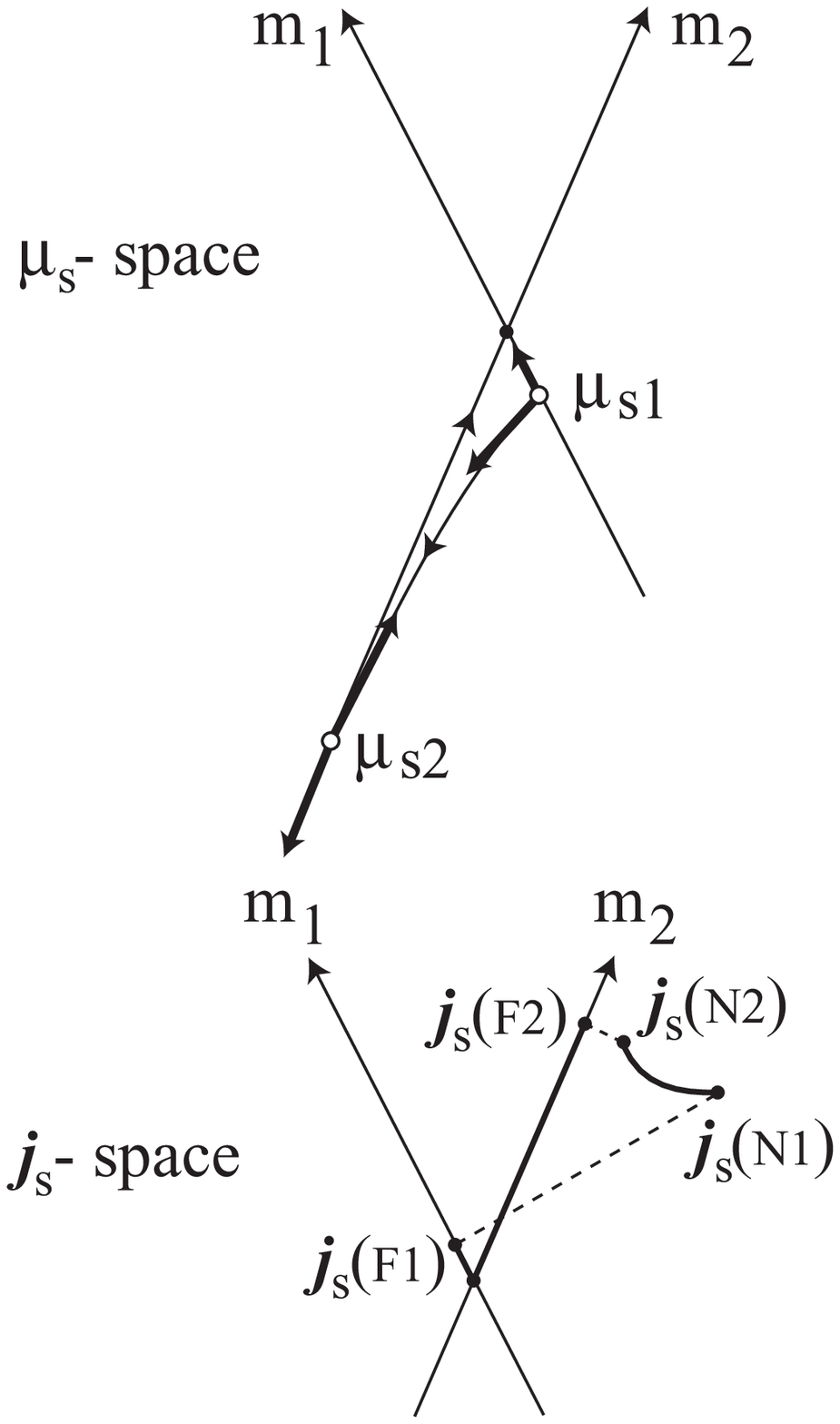}
  \caption{Hodographs of $\boldsymbol{\mu}_s(x)$ and ${\bf j}_s(x)$ in a situation with $g_2 < 0$.}
  \label{fig:parametric_unconventional}
\end{figure}

Directions of the ${\bf T}_1$ and ${\bf T}_2$ torques are determined by the spin current jumps, which in turn depend on the positions of the points ${\bf j}_s(N_1)$ and ${\bf j}_s(N_2)$ relative to the lines defined by the ${\bf m}_{1,2}$ vectors in Fig.\ref{fig:parametric_normal} (lower panel). The situation shown in Fig.\ref{fig:parametric_normal} corresponds to the positive values of $g_{1}$ and $g_2$. However, the efficiency factors signs can change if the material parameters of the ferromagnets are sufficiently different, i.e., in an asymmetric structure. From the point of view of mechanical ana\-logy the asymmetric structure is characterized by non-equal driving ($p_1 j_0 {\bf m}_1$ and $-p_2 j_0 {\bf m}_2$) and restoring ($- \tilde G_1 \boldsymbol{\mu}_{1}$ and $- \tilde G_2 \boldsymbol{\mu}_{2}$) forces. An example is shown in Fig.~\ref{fig:parametric_unconventional} (upper panel). Here the force $-p_2 j_0 {\bf m}_2$ pulls the N-cord with such strength that the $\boldsymbol{\mu}_{1}$ ring has been dragged to the negative side of the  ${\bf m}_1$ axis. Concurrent with this, the orientation of the hodograph's tangent at $\boldsymbol{\mu}_{2}$ has also changed: now the tangent line goes below the ${\bf m}_2$ axis. As the lower panel of Fig.~\ref{fig:parametric_unconventional} shows, this has led to a switch of the direction of spin current jump $\Delta {\bf j}_{s2}$. The sign of the torque ${\bf T}_2$ has switched together with it: instead of pulling ${\bf m}_2$ towards ${\bf m}_1$ it now pushes it away from ${\bf m}_1$. In other words, the system is characterized by $g_2 < 0$.

Transition from positive to negative $g_2$ happens when the ring $\boldsymbol{\mu}_{1}$ passes the crossing point of the rods. Note that the mutual orientation of the ${\bf j}_s(N_1)$ vector and ${\bf m}_1$ axis does not change during the transition, so $g_1$ remains positive.

In the example in Fig.~\ref{fig:parametric_unconventional} the $\boldsymbol{\mu}_{1}$ ring was dragged through the rods crossing poing because the $-p_2 j_0 {\bf m}_2$ force was increased. The same effect can happen due to a diminishing restoring force if the value of $\tilde G_2$ is sufficiently decreased.
With a weaker F$_2$-cord the $\boldsymbol{\mu}_{2}$ ring will slide further away from the origin along ${\bf m}_2$, and similarly drag the $\boldsymbol{\mu}_{1}$ ring through the origin.

\subsection{Singular limits of the spin-valve magnetoresistance}
\label{sec:MR}

\begin{figure}[b]
\includegraphics[width=0.4 \textwidth]{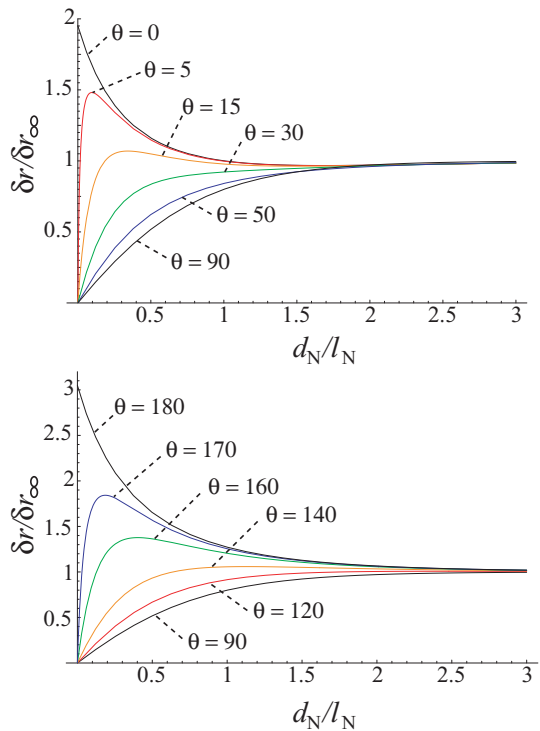}
\caption{Asymmetric F/N/F valve spin resistance $\delta r(d_N)$ relative to the resistance $\delta r_{\infty}$ at $d_N \to \infty$. Curves are marked by the values of the angle $\theta$ between the magnetizations of the ferromagnets. For clarity, angles between 0 and 90$^{\circ}$ and between 90$^{\circ}$ and 180$^{\circ}$ are shown in separate panels. Other parameters are set to $d_{i}/l_{i} \gg 1$ in ferromagnets, $G_1 = G_2$, $G_N = 4 G_1$, $p_1 = 0.1$, $p_2 = 0.9$}
  \label{fig:MR}
\end{figure}

We now turn to another example of mechanical analo\-gy lending a helping hand in understanding a seemingly paradoxical behavior of a layered spintronic device.

As discussed in the Introduction, the resistance of an F/N/F structure depends on the relative orientation of vectors ${\bf m}_1$ and ${\bf m}_2$. In spin valves that are engineered so that the angle $\theta$ between the magnetizations is controlled by external magnetic field one observes spin-related magnetoresistance (MR). Just as the spin torque, spin MR is produced by spin accumulation near the F/N boundaries. Such accumulation contributes to the total resistance even for a single F/N interface.\cite{vanson:1987} In an F/N/F valve spin accumulations at the two interfaces are not independent as long as the thickness of the N-layer is not much larger than the spin diffusion length. Because of that, the total extra resistance is magnetic configuration dependent.

A convenient way of thinking about spin MR is provided by a realization \cite{rashba_prbrc2000, rashba_epjb2002, fabian_APS2007, bazaliy:apl2017} that in ferromagnetic regions the gradient of spin potential acts as a distributed electromotive force (EMF) ${\mathcal E}(x) = - p \nabla \mu_s/2e$ (no EMF is generated in the normal parts of the structure). In a spin valve this EMF is directed against the flow of electric current, and thus increases the voltage needed to pump a given amount of current through the device. The full voltage increase is obtained by integrating the EMF
$$
\delta V = -\int_{-\infty}^{+\infty} {\mathcal E} dx
=  \frac{p_1 \mu_{s}(x_1) - p_2 \mu_{s}(x_2)}{2e} \ .
$$
Spin accumulations $\mu_s(x_{1,2})$ at the interfaces are found in Appendix~\ref{appendix-I}. They are proportional to $j_0$, so the effect of EMF can be indeed described by an extra resistance $\delta r$. The dependencies of this resistance on the thickness of the normal layer $d_N$ and the angle $\theta$ are shown in Fig.~\ref{fig:MR} for a particular asymmetric valve (MR of such valves was originally discussed in Ref.~\onlinecite{manschot:2004}).

A striking feature of the $\delta r(d_N)$ curves is that for all angles, except $\theta = 0$ and $\theta = 180^{\circ}$, the resistance vanishes at $d_N \to 0$. For the two special angles $\delta r$ approaches a finite value. Mathematically speaking, $\theta = 0$ and $\theta = 180^{\circ}$ are singular limiting points, where the value of $\delta r$ depends on the order of limits $d_N \to 0$ and $\theta \to 0, 180^{\circ}$

The vanishing of $\delta r$ looks puzzling from the physics point of view since at $d_N \ll l_N$ the N-layer can be considered practically free of spin relaxation. Spins are injected into a normal reservoir with no relaxation, and as the size of this reservoir get progressively smaller the spin accumulation vanishes. Why is it difficult to fill a small reservoir with spins? And why it becomes possible when the two magnetizations are exactly collinear?

\begin{figure}[t]
\includegraphics[width=0.45 \textwidth]{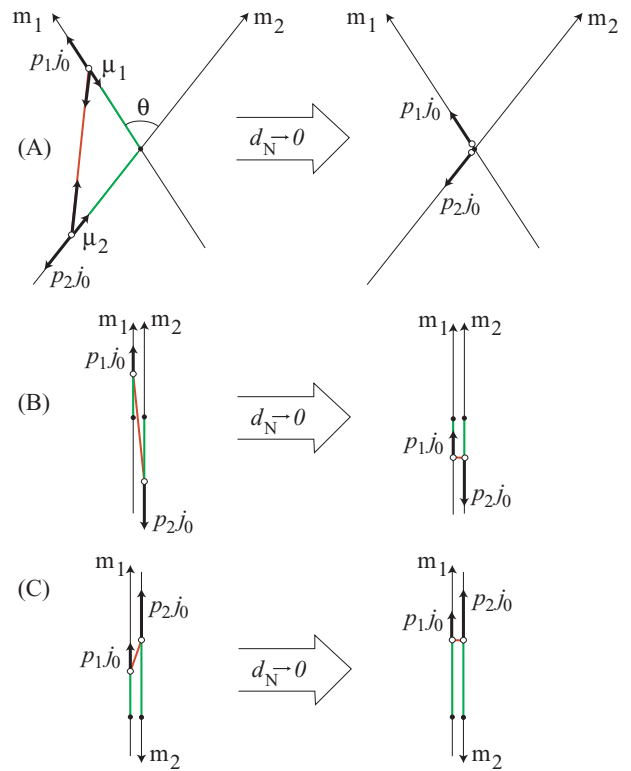}
\caption{Mechanical analogy for an F/N/F valve for (A) generic angle between magnetizations, (B) parallel configuration, (C) anti-parallel configuration. In all cases the situation is shown for finite $d_N \ll l_N$ on the left and for $d_N \to 0$ on the right. Red lines show the N-layer hodographs, green lines show the F-layer hodographs.}
  \label{fig:mech_small_dn}
\end{figure}

Let us first turn to the mechanical analogy. Fig.~\ref{fig:mech_small_dn} shows the elastic cords configurations for a generic angle $\theta$, and for the special cases of $\theta = 0$ and $\theta = 180^{\circ}$. In the $d_N \ll l_N$ limit the N-layer cord goes straight from one ring to another, as discussed in Sec.~\ref{sec:noncollinear}. The most important property of the $d_N \to 0$ limit is that the stiffness of the N-layer cord grows as $\tilde G_N(d_N) \sim 1/d_N \to \infty$ (Appendix~\ref{appendix-I}). The divergence of $\tilde G_N$ could have been expected even without calculations, since any changes happening over a progressively smaller distance inevitably produce infinite gradients. With an infinite stiffness of the N-layer cord, the distance between the two rings has to reduce to zero. It is evident from Fig.~\ref{fig:mech_small_dn}(A) that for a generic $\theta$ the only way to achieve this is to have both rings at the origin, i.e., have vanishing spin accumulations. The cases of $\theta = 0$ and $\theta = 180^{\circ}$ shown in Fig.~\ref{fig:mech_small_dn}(B,C) are exceptional: in collinear situations the rings can be together, yet not at the origin. Thus the N-layer cord can have zero length, and at the same time the driving forces can displace the rings away from the origin until a balance with the restoring forces of the F-layer cords is achieved. Note that for $\theta = 0$ the driving forces act in the opposite directions, so in the absence of asymmetry the $d_N \to 0$ positions of the rings will be at the origin. Overall, the mechanical analogy makes the behavior of the graphs in Fig.~\ref{fig:MR} quite obvious.

Armed with the insight from the mechanical analogy, we can understand this result in terms of spin physics. The decreasing thickness of the N-layer makes spin relaxation essentially negligible. Any difference between ${\boldsymbol \mu}_s(N_1)$ and ${\boldsymbol \mu}_s(N_2)$ produces a physically impossible infinite spin current. Therefore, the spin accumulation is required to remain constant throughout the N-layer. For $\theta = 0$ or $\theta = 180^{\circ}$ such requirement is fully compatible with the requirement of vector ${\boldsymbol \mu}_s$ being collinear with ${\bf m}_{1,2}$ at the boundaries with strong ferromagnets. But when ${\bf m}_{1}$ and ${\bf m}_{2}$ have a finite angle between them, the two requirements start to compete. Since in our model the ferromagnets are assumed to be so strong that the second requirement cannot be violated, the only compromise is to have zero ${\boldsymbol \mu}_s$ (which is formally collinear with both ${\bf m}_1$ and ${\bf m}_2$).

\section{Discussion}
The mechanical analogy allows one to qualitatively predict the behavior of layered diffusive devices. For example, it makes it evident that $g_2$ cannot become negative when the angle $\theta$ between the vectors ${\bf m}_1$ and ${\bf m}_2$ exceeds  90$^{\circ}$. Indeed, for such angles no driving force---regardless of how strong---acting on the $\boldsymbol{\mu}_{2}$ ring in the $-{\bf m}_2$ direction can drag the $\boldsymbol{\mu}_{1}$ ring to the negative side of the axis simply because  the component of such force along  ${\bf m}_1$ is positive (assuming, for definiteness, $j_0 > 0$). The sign change of  $g_2$ is therefore possible only for $\theta < \pi/2$. This conclusion formally follows from the expressions obtained in Refs.~\onlinecite{kovalev:2002, barnas:2005} (see Appendix \ref{appendix-II}), but with mechanical analogy we can make it without calculations. In the same spirit the mechanical analogy makes it evident that a sufficiently large polarization ratio $p_1/p_2$ will drag the $\boldsymbol{\mu}_2$ ring through the origin and cause the efficiency factor $g_1$ to change its sign, while $g_2$ will remain positive.

Furthermore, one can clearly see why both efficiency factors $g_1$ and $g_2$ cannot change signs simultaneously (Appendix~\ref{appendix-II}). The sign of $g$ changes when the corresponding ring moves through the origin due to the increasing asymmetry of the structure. Since both external forces pull the rings away from the rod crossing, at least one of the rings has to move in the direction of external force. The second ring may get to the origin if one force overpowers the other but there is no option for both rings to move towards the origin.\cite{negative_p}

Our last example in Sec.~\ref{sec:MR} shows that the me\-cha\-ni\-cal analogy may provide critical insights into spintronic  problems even before the actual spin physics is understood on a qualitative level.

Is it possible to use the mechanical analogy for structures with more than two ferromagnetic layers? In such systems the hodograph $\boldsymbol{\mu}_s(x)$ in each N$_i$-layer is a 2D curve in the spin space that belongs to a plane defined by the magnetizations ${\bf m}_{i-1}$ and ${\bf m}_{i+1}$ of the F$_{i-1}$ and F$_{i-1}$-layers sandwiching the N$_i$-layer. In the F-layers the hodographs are straight lines along the direction of $\bf m$. Magnetizations of the F-layers may be non-coplanar, so every N-layer will have its own 2D plane of the hodograph. However, this complication does not prevent one from showing all hodographs together on the diagrams similar to those in Figs.~\ref{fig:parametric_normal} and \ref{fig:parametric_unconventional}. Indeed, in the nearby N$_{k-1}$ and N$_{k+1}$-layers, separated by the F$_k$-layer, the 2D planes have a common line, the ${\bf m}_k$ axis. For that reason the planes forming the full 3D hodograph can be sequentially folded into a stack of sheets, one on top of the other. After such folding, the hodograph will become a collection of N-curves going between the ${\bf m}_1$,  ${\bf m}_2$ \ldots ${\bf m}_n$ axes.

The mechanical analogy is easily extended to the multi-layer wires. Every boundary is represented by a rod with a weightless ring sliding along it. The balance of driving and restoring forces determines the position of each ring. With the increasing number of rings and rods it becomes progressively more difficult to determine the equilibrium state of the system based on the mechanical intuition alone. Nevertheless, our graphic interpretation can still serve as a useful tool for presenting the analytic or numeric results. Moreover, in the case of small changes of parameters the mechanical analogy will often provide an easy way to predict the system's response.

\section*{Acknowledgements}
This work was supported by the NSF DMR-0847159 grant.

\appendix

\section{Spin accumulation and spin current distributions}\label{appendix-I}
In this Appendix we provide the results of spin accumulation calculations for a N/F$_1$/N/F$_2$/N structure with the ferromagnetic layers having thicknesses $d_{1,2}$ and the normal layer having a thickness  $d_N$. The spin diffusion lengths in the ferromagnets are denoted as $l_{1}$ and $l_{2}$. Spin diffusion length in the central normal layer is $l_{N}$. We assume vanishing spin diffusion lengths in the two outer N-layers, thus spin accumulations at the outer N/F boundaries vanish as well. Spin accumulations at the F$_1$/N ($x = x_1$) the N/F$_2$ ($x = x_2$) boundaries are denoted as $\boldsymbol{\mu}_1$ and $\boldsymbol{\mu}_2$. The function $\boldsymbol{\mu}(x)$ is given buy the following expressions. In F$_1$ ($x_1 - d_1 < x < x_1$)
$$
\boldsymbol{\mu}(x) = \frac{\boldsymbol{\mu}_1 \sinh [(x-(x_1-d_1))/l_1]}{\sinh d_1/l_{1}  } \ .
$$
In N ($x_1 < x < x_2$)
$$
\boldsymbol{\mu}(x) = \frac{\boldsymbol{\mu}_1 \sinh [(x_2-x)/l_N]
 + \boldsymbol{\mu}_2 \sinh [(x-x_1)/l_N]}{\sinh d_N/l_{N}  }  \ .
$$
And in F$_2$ ($x_2 < x < x_2 + d_2$)
$$
\boldsymbol{\mu}(x) = \frac{\boldsymbol{\mu}_2 \sinh [((x_2+d_2)-x)/l_2]}{\sinh d_2/l_{2}  }
$$
Denoting $\tilde G(d) = G \coth (d/l)$ we get the following expressions for spin currents at the F$_1$/N and N/F$_2$ interfaces
\begin{eqnarray}
 \nonumber
{\bf j}_s(F_1) &=& p_1 j_0 - \tilde G_1(d_1) \boldsymbol{\mu}_1 \ ,
 \\
  \nonumber
{\bf j}_s(N_1) &=& \tilde G_N(d_N) \left(\boldsymbol{\mu}_1
 - \frac{1}{\cosh d_N/l_{N}} \boldsymbol{\mu}_2 \right) \ ,
 \\ \label{appI:js}
{\bf j}_s(N_2) &=& \tilde G_N(d_N) \left(\frac{1}{\cosh d_N/l_{N}} \boldsymbol{\mu}_1
 -  \boldsymbol{\mu}_2 \right) \ ,
 \\
  \nonumber
{\bf j}_s(F_2) &=& p_2 j_0 +  \tilde G_2(d_2) \boldsymbol{\mu}_2 \ ,
\end{eqnarray}
Using (\ref{appI:js}) in the boundary conditions (\ref{BC-Jparallel}) we then find an expression for spin accumulations at the boundaries
\begin{equation}\label{appI:mu1mu2}
\left(\begin{array}{c} \mu_1 \\ \mu_2 \end{array}\right) =
 \frac{j_0}{\rm Det}
 \left| \begin{array}{cc}
 A_2 & -B\cos\theta \\
 B\cos\theta & -A_1
 \end{array}\right|
 \left(\begin{array}{c} p_1 \\ p_2 \end{array}\right) \ ,
\end{equation}
where
\begin{eqnarray*}
A_1 &=& \tilde G_1(d_1) + \tilde G_N(d_N) \ ,
 \\
A_2 &=& \tilde G_2(d_2) + \tilde G_N(d_N) \ ,
 \\
B &=& \frac{\tilde G_N(d_N)}{\cosh(d_N/l_N)} \ ,
 \\
{\rm Det} &=& A_1 A_2 - B^2 \cos^2\theta \ .
\end{eqnarray*}

\section{Spin-transfer torques}\label{appendix-II}
The jumps of spin currents at the interfaces are given by formulas (\ref{eq:js_jumps}). Substituting expressions (\ref{appI:js}) into them we get
\begin{eqnarray*}
\Delta {\bf j}_{s1} &=& \frac{\tilde G_N(d_N)}{\cosh(d_N/l_N)} \ \mu_2 \
  [{\bf m}_1 \times [{\bf m}_2 \times {\bf m}_1]] \ ,
 \\
\Delta {\bf j}_{s2} &=& \frac{\tilde G_N(d_N)}{\cosh(d_N/l_N)} \ \mu_1 \
   [{\bf m}_2 \times [{\bf m}_1 \times {\bf m}_2]]  \ ,
\end{eqnarray*}
where $\mu_{1,2}$ have to be substituted from (\ref{appI:mu1mu2}).

To analyze the signs of the efficiency factors one first proves the inequalities $0< B < A_{1,2}$. Then
it is easy to show that the signs of $g_{1,2}$ switch from positive to negative only if $\cos\theta > 0$, and one of the following conditions is satisfied
\begin{eqnarray}
\nonumber
&& \frac{p_2}{p_1} > \frac{A_2}{B\cos\theta} > \frac{A_2}{B} > 1 \quad \Rightarrow \quad g_2(\theta) < 0 \ ,
 \\
 \label{eq:switching_conditions}
&& \frac{p_2}{p_1} < \frac{B\cos\theta}{A_1} < \frac{B}{A_1} < 1 \quad \Rightarrow \quad g_1(\theta) < 0 \ .
\end{eqnarray}
Both conditions (\ref{eq:switching_conditions}) cannot be satisfied at the same time, so at most one efficiency factor can be negative for a given angle $\theta$ and device parameters. For $p_2/p_1 > A_2/B$ efficiency factor $g_2(\theta)$ is sign-changing as a function of $\theta$. For $p_2/p_1 < B/A_1$ efficiency factor $g_1(\theta)$ is sign-changing as a function of $\theta$. In both cases $g$ becomes a sign-changing function after the spin valve asymmetry exceeds a certain threshold.


\begin{thebibliography}{2}
\bibitem{pipe-analogy}
The logic here is similar to that in the case of a rubber tube with water flowing through it under pressure. From the fact that it is easier for the water to flow through a tube of a larger diameter we can conclude that pumping water will cause the tube to expand its cross section. This argument, and the one in the text, are examples of the Le Chatelier's principle.
\bibitem{negative_p}
One may ask if these arguments could change for $p < 0$ ($D_{\uparrow} < D_{\downarrow}$) in one or both F-layers. Those cases, however, can be fully accommodated by inverting the $\bf m$ vector in the layer with negative transport polarization.
\end{thebibliography}
\end{document}